\documentclass[12pt]{article}

\usepackage{scicite}
\usepackage{times}
\usepackage{graphicx}
\usepackage{float}

\usepackage{caption} 
\usepackage{soul}
\usepackage{color}							

\definecolor{mLink}{rgb}{0.0,0.0,0.0}				
\definecolor{mCite}{rgb}{0.0,0.0,0.0}	
\definecolor{mURLs}{rgb}{0.0,0.0,0.0}

\DeclareCaptionLabelSeparator{lsep}{ | }
\newenvironment{sciabstract}{%
\begin{quote} \bf}
{\end{quote}}

\title{Conflict and Convention in Dynamic Networks}

\author
{Michael Foley,$^{1\ast}$ Patrick Forber,$^{2}$ Rory Smead,$^{3}$ Christoph Riedl$^{456\ast}$\\
\\
\normalsize{$^{1}$Network Science Institute, Northeastern University, Boston, Massachusetts, USA}\\
\normalsize{$^{2}$Department of Philosophy, Tufts University, Medford, Massachusetts, USA}\\
\normalsize{$^{3}$Department of Philosophy and Religion, Northeastern University, Boston, Massachusetts, USA}\\
\normalsize{$^{4}$D'€™Amore-McKim School of Business, Northeastern University, Boston, Massachusetts, USA}\\
\normalsize{$^{5}$College of Computer and Information Science, Northeastern University, Massachusetts, USA}\\
\normalsize{$^{6}$Institute for Quantitative Social Science, Harvard University, Cambridge, Massachusetts, USA}\\
\\
\normalsize{$^\ast$To whom correspondence should be addressed; E-mail:  c.riedl@neu.edu}
}

\date{}

\usepackage[					
        pdftex,
        colorlinks=true,
        citecolor=mCite,
        linkcolor=mLink,
        filecolor=magenta,
        urlcolor=mURLs
        ]{hyperref}

\begin{document} 




\maketitle

\begin{sciabstract}
An important way to resolve games of conflict (snowdrift, hawk-dove, chicken) involves adopting a convention: a correlated equilibrium that avoids any conflict between aggressive strategies. Dynamic networks allow individuals to resolve conflict via their network connections rather than changing their strategy. Exploring how behavioral strategies coevolve with social networks reveals new dynamics that can help explain the origins and robustness of conventions. Here we model the emergence of conventions as correlated equilibria in dynamic networks. Our results show that networks have the tendency to break the symmetry between the two conventional solutions in a strongly biased way. Rather than the correlated equilibrium associated with ownership norms (play aggressive at home, not away), we usually see the opposite host-guest norm (play aggressive away, not at home) evolve on dynamic networks, a phenomenon common to human interaction. We also show that learning to avoid conflict can produce realistic network structures in a way different than preferential attachment models.
\end{sciabstract}

\section{Introduction}
An essential feature of cooperation involves resolving conflicts, and an effective way to resolve conflicts is to adopt a convention that avoids them altogether.  Evolutionary game theory provides a way to represent the conflicts associated with the problem of cooperation \cite{fudenberg1991game,weibull1997evolutionary,nowak06fiverules}, and reinforcement learning can produce adaptive solutions to broad classes of strategic interactions \cite{herrnstein1970law,roth1995learning,erev1998predicting,fudenberg1998theory,macy-flache2002,beggs2005convergence,huttegger2014-pnas}. In games of conflict (snowdrift, hawk-dove, chicken) \cite{weibull1997evolutionary} individuals can learn to a play a {\em correlated equilibrium}, similar to turn-taking, that does better than the mixed Nash equilibrium \cite{aumann1974subjectivity,smith1982evolution,skyrms2014evolution,kakade2003correlated}. These solutions, often understood as an important aspect of conventions, play a central role in evolutionary accounts of coordinated action, territoriality, and ownership \cite{smith-parker1976,smith1982evolution,young93,vanderschraaf95,kokko-etal06,alexander07,guala-mittone10,mesterton-gibbons-etal14,skyrms2014evolution}. However, most studies of convention examine learning or evolution with random interactions or static network structures. In reality, learning affects social networks as well as behavior. Individuals may seek out or avoid others based on past interactions and this can have profound effects on strategy choices, as work on dynamic networks has shown \cite{jackson2002formation,goyal2005,santos2005scale,ohtsuki2006simple,pacheco2006coevolution,szabo07,gross08,perc10,wuetal10,rand2011dynamic,pinheiro12,pinheiro16,allen-etal17,Wu2010dynamicNetworks}, an effect also seen in empirical work on human behavior \cite{FehlEtAl2011coevolution}. For games of conflict, dynamic networks allow individuals to avoid aggressive neighbors by choosing their interaction partners which tends to promote cooperation \cite{santos2006cooperation}, although some static network structures work against cooperation \cite{hauert2004spatial}. 

Here we present novel results regarding the emergence of conventions as correlated equilibrium solutions for games of conflict. We use a coevolutionary model where both network ties and behavioral strategies evolve by reinforcement learning. Few previous studies model co-evolution of network and strategy evolving with reinforcement learning \cite{skyrms2000dynamic,pemantle-skyrms2004,alexander07} and none have explored how dynamic networks influence the evolution of correlated equilibria. We employ computer simulations to search the entire payoff space for games of conflict and to explore dynamic behavior over a wide range of additional model parameters, including population size. We find that correlated equilibria often emerge spontaneously as a natural result of reinforcement learning on networks, and that dynamic networks strongly favor the so-called paradoxical solution where individuals act aggressively away but not at home. This result suggests that the prevalence of territoriality and ownership behaviors is not primarily a matter of convention. Our results also have relevance to general mechanisms of network formation. Learning to avoid conflict by evolving network ties can produce realistic network structures and thus provides another potential mechanism for social network formation that differs from preferential attachment \cite{BarabasiAlbert1999emergence,pacheco2006coevolution,santos2006cooperation}. 

In games of conflict each agent can play one of two strategies: an aggressive, competitive strategy (hawk) or a more passive, cooperative strategy (dove). For economy of presentation we will use the hawk-dove terminology but the results apply equally to any game of conflict meeting the payoffs defined in Fig.~\ref{fig:Panel1}a \cite{weibull1997evolutionary}. The highest payoff results from playing hawk against dove, the lowest from playing hawk against hawk and all players would prefer their opponent to play dove. Agents playing these games either coordinate on who gets the large payoff, compete for a resource, or engage in a standoff. In an unstructured population common evolutionary dynamics (e.g., replicator dynamics) converge on the mixed Nash equilibrium of the game: host plays dove with probability $x_1/(1-y_1+x_1)$ and the visitor plays dove with probability $x_2/(1-y_2+x_2)$. 

Introducing a dynamic network permits individuals to find solutions different than the mixed Nash (summarized in Fig.~\ref{fig:Panel1}c). If network ties are updated symmetrically using reinforcement learning (i.e., both players update their network ties based on their interaction payoffs as in \cite{skyrms2000dynamic,pemantle-skyrms2004}), there is a new potential solution to games of conflict: agents can learn pure strategies while learning to interact with suitable partners. In these cases a hub-and-spoke network structure tends to emerge. One agent becomes a hub and learns to play a pure dove strategy whereas other agents in the population learn to play pure hawk and only visit the dove hub. In effect, agents structure their interaction network to avoid conflict, allowing for a kind of cooperation in games of conflict, as previous work has shown \cite{santos2006cooperation}.

Another potential new solution, in the form of efficient correlated equilibria, can emerge when individuals are able to coordinate their strategies on an external cue or signal. Correlating behavior on some external cue (say a coin toss) can produce a {\em convention}, such as taking turns playing aggressively (e.g., player one plays hawk when heads, dove otherwise; and player two plays hawk when tails, dove otherwise) \cite{aumann1974subjectivity,skyrms2014evolution}. Indeed, correlated equilibria can be learned by relatively simple learning rules \cite{foster-vohara1997correlated,hart-mas-colell2000correlated,arifovic-etal2016correlated}, and are also readily learned by humans \cite{duffy-feltovich-2010correlated}. In the context of networks, correlated equilibria may arise if individuals learn their host and visitor behavior independently. This enables them to play one way when hosting and another way when visiting, using their role as host versus visitor as the public cue (Fig.~\ref{fig:Panel1}b). At one such correlated equilibrium individuals play what is called the {\em bourgeois} strategy, where they play hawk when hosting and dove when visiting. This strategy has been invoked as an explanation for norms of ownership or territoriality \cite{smith-parker1976,smith1982evolution}. That is, norms of ownership may just be a conventional solution to games of conflict, rather than reflecting the intrinsic value of the resource or territory. However, there is another equally efficient solution that uses the {\em paradoxical} strategy, where one plays dove when hosting and hawk when visiting. This alternative solution complicates the evolutionary account of ownership and territoriality. In unstructured populations with symmetric payoffs both solutions are strongly stable and evolve with equal frequency \cite{skyrms2014evolution}. This presents a problem of symmetry-breaking: why does it seem that the bourgeois strategy is so much more prominent in nature (see, e.g., \cite{krebs-davies1993}) when there is another equally viable option? Previous work has argued that individual learners who are sensitive to relatively small inequalities in value of the territory or fighting ability of the current owner may break the symmetry and strongly favor the bourgeois solution \cite{skyrms2014evolution}. Here we develop a model to investigate the emergence and stability of these correlated equilibria for games of conflict. 


\section{Methods}


\subsection{The Model}

\begin{figure*}[ht!]
\centering
\includegraphics[width=\linewidth]{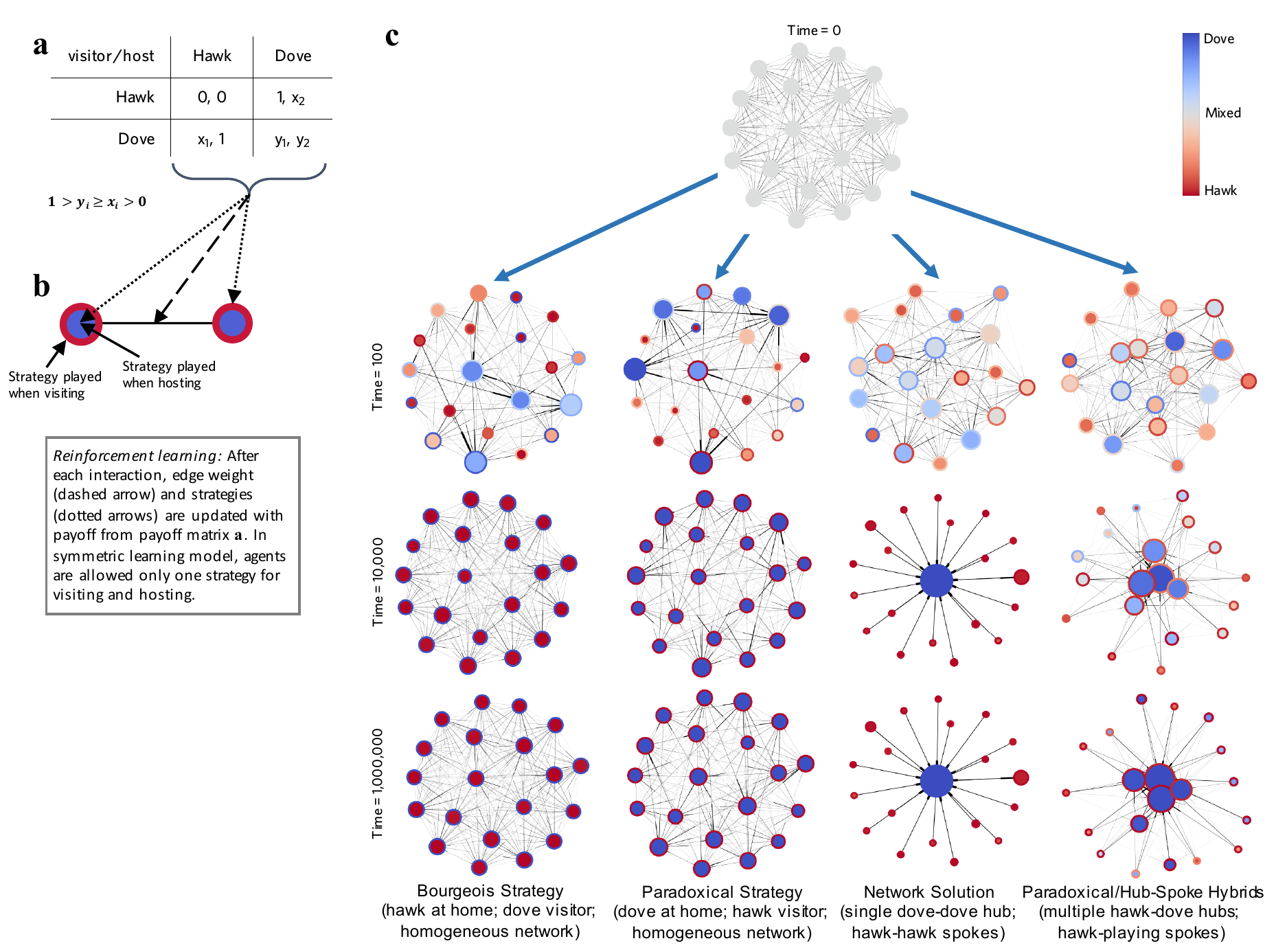}
\caption{\textbf{Coevolution of network ties and strategies.}
\textbf{a}, Payoff matrix for the hawk-dove game. 
\textbf{b}, Reinforcement learning for host and visitor roles.
\textbf{c}, Starting from uniform homogeneous mixing (fully connected network) and random strategies, populations tend to evolve into one of four stable outcomes depending on early random events and simulation parameters: Bourgeois, Paradoxical, Network or Hybrid solutions. Node colors represent strategy weights, node size is proportional to number of expected visitors and stronger connections are denoted with darker lines. 
}
\label{fig:Panel1}
\end{figure*}

In our dynamic network model agents simultaneously learn how to behave and with whom to interact via Roth-Erev reinforcement. Roth-Erev reinforcement learning provides a simple yet psychologically realistic learning dynamic in which an individual's behavior evolves iteratively in response to the payoffs obtained through interaction \cite{erev1998predicting}. This contrasts with other models of evolutionary game theory, such as the replicator dynamics, where ``learning''  (evolution) occurs globally across the populations as individuals reproduce at rates proportional to the success of their behavioral strategies. In reinforcement learning, individuals have weights associated with each possible behavior and these weights are updated based on the payoffs received during their interactions (Fig.~\ref{fig:Panel1}a--b). Unlike imitation learning, reinforcement learning does not require individuals to track the behaviors or payoffs of other players in the population, a common feature in contagion models \cite{dodds2005generalized}; an agent only tracks the payoffs associated with their own behavior and update their learning weights accordingly. Initial weights determine the initial behavior of the system. For instance, when initial weights are equal and relatively small we see effectively random behavior on a uniform network at the start. At every time step each agent visits another agent to play a game, receives a payoff, and updates learning weights. This update can be symmetric---both host and visitor update their network connection and strategy weights simultaneously---or asymmetric---the hosts only update their strategy choices when hosting whereas the visitors update both their network connection and strategy choices when visiting. The rationale for this particular asymmetry is that while individuals can control their behavioral strategies when hosting or visiting, and who they visit, they cannot control who decides to visit them. Specifically, we implement this asymmetry by having separate learning weights for hosting and visiting roles, allowing individuals the potential to learn different host and visitor strategies. Also, since hosts have no control over who decides to visit, only visitors update their network connections based on the interactions payoffs. 

The asymmetry in network updating allows for correlated equilibria to emerge and evolve with respect to player roles and therefore provides a way to investigate the conventional origins of norms of ownership or territoriality. Also, in contrast to many previous dynamic network models that allow network ties to form or dissolve discretely \cite{goyal2005,pacheco2006coevolution,santos2006cooperation,rand2011dynamic}, our model allows network connections to strengthen or weaken as a function of interaction payoffs. Since ties are rarely discrete in social networks \cite{Granovetter:1973strength,barrat2004weightedPNAS,yook2001weighted}, modeling network connections with reinforcement learning adds an important element of realism. Furthermore, modeling network ties through reinforcement learning avoids the need to fix the number of network ties which can affect the results \cite{rand2014static}. Instead, we allow the number (strength) of ties to evolve endogenously in response to learning.

More precisely, we model a set of $N$ agents engaged in pairwise games of conflict across a number rounds. Each round, every agent choses one other agent to visit and engages in an interaction. During each interaction the host and visitor independently choose a behavior, receive a payoff, and learn via reinforcement. All players have one interaction as a visitor and up to $N-1$ interactions as a host.

Each agent $i$ has two vectors $(w_H, w_D)$ and $(w_h, w_d)$ with $i$'s reinforcement weights for each strategy when hosting or visiting respectively. Each agent $i$ also has a vector representing their reinforcement weights for choice of which player to visit: $(w_{i1}, w_{i2},...,w_{in})$ where $w_{ij}$ represents the weight related to player $i$ visiting player $j$. Self-visits are not allowed $(w_{ii} = 0)$. Initial weights for both strategy choices are set to 1, while network partner weights are uniformly set to $\frac{L}{N-1}$. We adopted the convention of $L=19$ so network partner weights for the smallest population size we study ($N=20$) would start at 1 ($w_{ij} = 1$ for all $i,j$ with $i\neq j$ as the baseline). We kept $L$ constant across different population sizes to ensure that the reinforcement learning has a similar speed relative to total initial weights in larger population sizes. Changing $L$ effectively varies network learning speed; larger $L$ values slow the responsiveness of an agent to interaction payoffs. (See Supplemental Information for further details about initial weights and varying learning speeds.)

The model has two modular dynamical components: discounting ($\delta$) and errors ($\epsilon$). Discounting involves reducing past learning weights as more reinforcement occurs, effectively allowing agents to forget older strategies and network connections \cite{erev1998predicting,pemantle-skyrms2004}. Errors represent mistakes, mutations, or noise where an agent selects a strategy or node in the network at random rather than according to their learning weights. Both of these components are known to impact the stability and long-run behavior of reinforcement learning \cite{skyrms2000dynamic}.

When taking an action, for a given agent in a given context (Host, Visitor, Partner Choice), the probability of choosing option $s$ is proportional to the current relevant weights: 

\begin{equation}
Pr(s) = (1-\epsilon) \frac{w_s}{\sum_{s'} w_{s'}} + \epsilon \frac{1}{|S|}
\end{equation}

\noindent where $\epsilon$ is the error rate, $S$ is the relevant set of available choices (i.e., visiting strategies, hosting strategies, or choice of player to visit), $s \in S$ and $s' \in S$.

After each instance of learning, the relevant weights are discounted by a factor $\delta$ and increased by the received payoff $(\pi)$.

\begin{equation}
w_s' = (1-\delta) w_s + \pi_s
\end{equation}

\noindent where $w_s'$ is the weights after updating and $\pi_s$ is the most recent payoff related to choosing action $s$. $\pi_s = 0$ if action $s$ was not chosen. All visits and strategy choices occur simultaneously within a round. All weights are updated simultaneously after each round at a rate of one update per related interaction from that round. Payoffs received for strategy choice when visiting do not alter the weights for strategy choice when hosting and vice versa. Likewise, payoffs received for hosting do not affect adjacency weights for choosing which other agents to visit.

\section{Results}

We explore our model through simulations of reinforcement learning on the asymmetric dynamic network. To provide some context for interpreting the simulation results, let us first describe the sort of solutions we see for games of conflict on dynamic networks. First, there are correlated equilibrium solutions. Note that a population may reach a correlated equilibrium by coordinating their strategies on whether they are host or visitor in one of two ways. Agents may learn to play hawk when hosting and dove when visiting. Hosts always receive the maximal payoff and visitors always receive $x_2$ (the bourgeois solution). Alternatively, agents may learn to play dove when hosting and hawk when visiting. This equilibrium grants the visitors the maximal payoff while the hosts receive $x_1$ (the paradoxical solution). In these cases the resulting network is a homogeneous, fully connected graph that approximates a randomly mixing population. Once the correlated equilibrium is established, visiting any agent yields the same payoff as everyone is deploying the same strategies conditional on the same cues. These correlated equilibrium solutions effectively avoid any conflict between aggressive hawk strategies, and these solutions are commonly found in asymmetric network runs with discounting and errors.

\begin{figure*}[ht!]
\centering
\includegraphics[width=.75\linewidth]{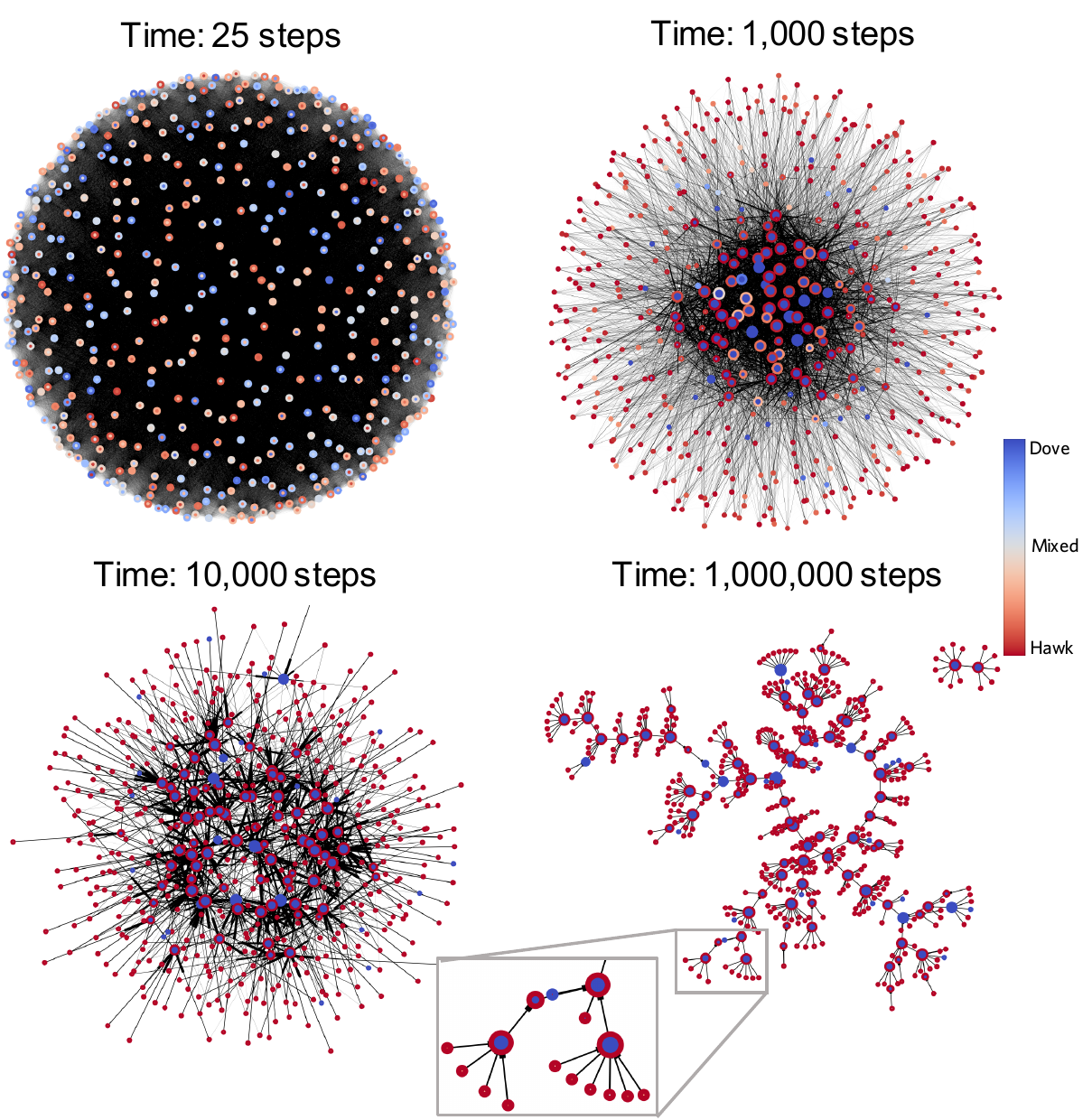}
\caption{\textbf{Illustrative example.} 
A representative result of coevolution of network and strategies in a larger population (N=500, discount $\delta=.01$, no error, payoff $x_1=x_2=0.2$, $y_1=y_2=0.6$). Without errors the network approaches a complicated hybrid solution. Node size is proportional to number of expected visitors.
}
\label{fig:Panel2}
\end{figure*}

Distinct from correlated equilibrium solutions, there are network solutions that mitigate conflict by utilizing network structure alone: if any agent starts to lean towards dove as host then agents learn to visit that dove-leaning agent more and also learn to play hawk when they visit. This produces a hub-spoke network structure with one dove-host attracting many hawk-visitors. In symmetric dynamic network runs we usually see network solutions emerge. Hybrid solutions involve a combination of the paradoxical correlated convention with the network solution where a small number of dove hubs learn the correlated equilibrium and visit each other while hawk spokes learn a pure (aggressive) strategy and maintain connection to the paradoxical dove-as-host hubs. Hybrid solutions usually emerge for asymmetric network runs with discounting but no errors (e.g., Fig.~\ref{fig:Panel2}). 

\begin{figure*}[]
\centering
\includegraphics[width=\linewidth]{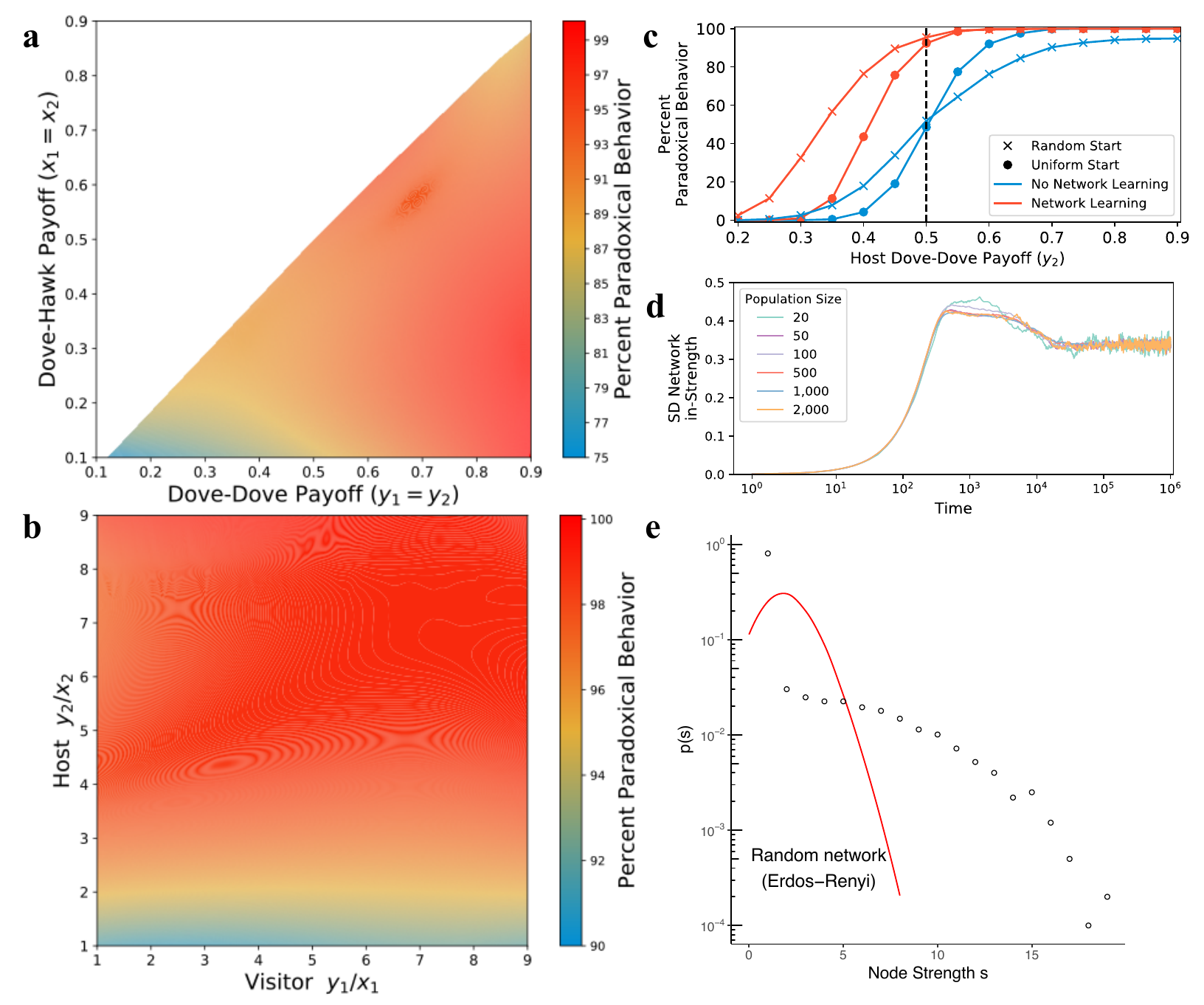}
\caption{\textbf{Simulation results.} 
\textbf{a} Heatmaps representing the proportion of outcomes approaching the paradoxical convention for the space of all possible games of conflict with symmetric payoffs (discount $\delta=.01$, error $\epsilon=.01$, N=20).
\textbf{b} Heatmap representing the proportion of outcomes approaching paradoxical convention for a slice of the asymmetric payoff space where $y_2 + x_2 = 1$, $y_1 + x_1 = 1$. The x axis shows the ratio of dove-dove to dove-hawk payoffs for visitors, while the y axis shows the same ratio for the hosts' payoffs. This ratio represents the relative incentive to find a dove parter while playing dove. ($\delta=.01$, $\epsilon=.01$, N=20).
\textbf{c} The proportion of paradoxical behavior produced by simulations runs (1,000 seeds) for random versus uniform initial learning weights both with and without network learning.  Payoffs are as follows: $y_1 = 0.5$, $x_1 = 0.4$, $y_2$ varies from 0.2 to 0.9 (along the x-axis), and $x_2 = y_2 - 0.1$.  The vertical dashed line represents symmetrical payoffs.  When $y_2 < y_1$, payoffs are biased towards the bourgeois strategy, and when $y_2 > y_1$, payoffs are biased towards the paradoxical strategy.
\textbf{d}, Convergence to homogeneous network in the Bourgeois and Paradoxical solution across different population sizes.  Parameters in this plot are as follows: Payoffs $x_2 = 0.3$, $y_2 = 0.5$, $x_1 = 0.6$, and $y_1 = 0.8$, $\delta = 0.01$, $\epsilon = 0.01$.
\textbf{e}, The degree distribution of networks across simulation results shows heterogeneity of network ties across individuals in the population ($\delta=.01$, no error, N=1,000, payoff $x_1=x_2=0.2$, $y_1=y_2=0.6$, average of 10 simulations shown).
}
\label{fig:Panel3}
\end{figure*}

Across all simulations we find that agents arrive at solutions to the games of conflict that fall into one of the four families described above: the two potential conventions (bourgeois and paradoxical), network solutions, or hybrid solutions (as in Fig.~\ref{fig:Panel1}c). We find no instances of the the mixed Nash equilibrium and simulation runs approach one of the four solution families by one million time steps with only very rare exceptions. The relative frequency of outcomes approaching each family varies with the presence of errors, the rate of discounting, and the specific values of the payoffs. Here we present the results with discounting for both network connections and strategies, as well as with and without errors. These results are qualitatively representative for a constellation of other factors, including larger population sizes, different relative learning speeds for strategy and networks connections, and learning with no discounting. Additional results are discussed in the Supplemental Information. 

We conducted an exhaustive search of the entire payoff space for games of conflict (omitting near boundary cases where $y_i$ or $x_i$ approach 0 or 1). As shown in Fig.~\ref{fig:Panel3}a, the paradoxical solution is by far the most prevalent outcome of simulation runs for symmetric payoffs. We also investigated the case where hosts and visitors have different payoff functions, effectively valuing resources differently when host versus visitor. When payoffs are biased in this way, such that it may be more valuable to play a particular strategy (hawk or dove) at home rather than away (or vice versa), the paradoxical solution is still prevalent, though less so when payoffs are strongly biased towards the bourgeois solution (Fig.~\ref{fig:Panel3}b; SI). It is important to note that biasing the payoffs in this way makes the bourgeois solution Pareto optimal and, in many contexts, makes it the preferred or expected equilibrium of evolutionary processes \cite{sandholm2010population,szabo-borsos2016potential}. The prevalence of the paradoxical solution in cases where the payoffs favor the bourgeois solution reveals the influence dynamic networks on breaking the symmetry of conventions.

Initializing simulations with random rather than uniform starting learning weights on initial strategies further increases the prevalence of paradoxical results on dynamic networks (Fig.~\ref{fig:Panel3}c). As a control case, we examined the model without network learning where agents simply choose whom to visit at random every round while still including asymmetric strategy updating. This is equivalent to a randomly mixing population. As one would expect in this case, we see an even split between bourgeois and paradoxical solutions when payoffs are symmetrical (Fig.~\ref{fig:Panel3}c). When payoffs are biased towards one solution or another we see a proportional increase in that solution evolving, with random learning weights introducing more noise. Perhaps surprisingly, the dynamical results are insensitive to population size---we see the same qualitative dynamics in populations of sizes from 20 to 2,000 (Fig.~\ref{fig:Panel3}d).

Without errors discounting leads to behavioral trapping; agents may eventually forget strategy choices and network connections, sometimes producing suboptimal results \cite{pemantle-skyrms2004}. In our model discounting without errors produces stable complex hybrid solutions with only rare exceptions. Fig.~\ref{fig:Panel2} shows the network evolution and emergence of a hybrid solution for a large population without errors. These networks have heterogeneous degree distributions that are different from random (i.e., Erdos-Renyi) networks (Fig.~\ref{fig:Panel3}e). The difference from random networks arises in the tail of the degree distribution, indicating a substantial probability to observe nodes with far more connections than the typical value. This reflects the formation of hubs (either pure dove or paradoxical) that attract many visitors, illustrating a phenomenon found in many real-world networks \cite{BarabasiAlbert1999emergence,clauset2009power,watts2007influentials}. Learning to avoid conflict via reinforcement learning without discounting thus produces a result relevant to current network science: we see realistic network structures produced by an alternative mechanism of social network formation that differs from the preferential attachment and fitness models. Those models are based on a mechanism of growing the network by adding nodes that preferentially attach to already highly-connected nodes in the network. Our social network formation model is a self-organizing mechanism that is based on a network with a fixed number of nodes in which tie strengths are modified to avoid conflict.

\begin{figure*}[]
\centering
\includegraphics[width=\linewidth]{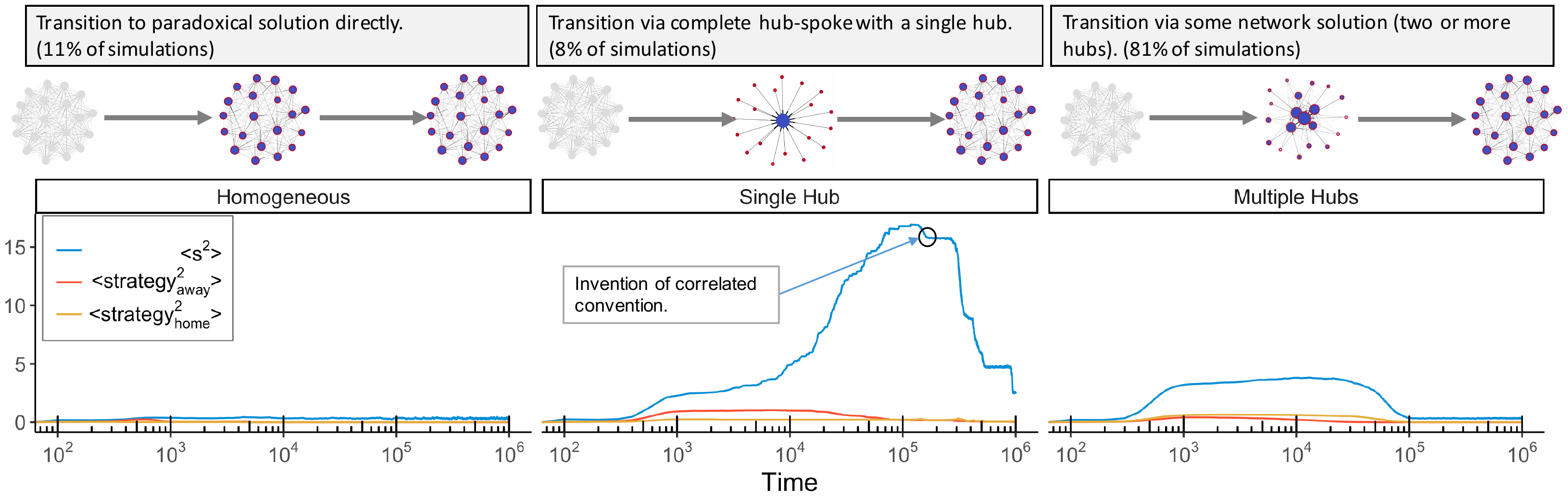}
\caption{\textbf{Temporal evolution.} 
Analysis of the possible evolutionary trajectories that produce the paradoxical solution. While some populations find the paradoxical solution right away, most transition to the paradoxical convention by first developing a hub-spoke structure indicative of network or hybrid solutions.
Plots show mean value across all simulations that fall into a specific evolutionary path. Node size shown proportional to number of expected visitors.
}
\label{fig:Panel4}
\end{figure*}

Errors promote exploration and allow learning to avoid the trapping states. When including an error rate ($\epsilon = 0.001$ or $0.01$) the system becomes ergodic, but simulations in the medium-run always show populations close to one of the four solution families. Most often the system is found near one of the conventional solutions with a near homogeneous network. In contrast to Fig.~\ref{fig:Panel2}, the inclusion of errors tends to eliminate the dove hubs found in hybrid solutions. Through exploration allowed by errors, the dove hubs tend to find each other and quickly learn the paradoxical solution. Once this starts to catch on, the rest of the population follows, which produces a more homogeneous network playing a correlated equilibrium (see Fig.~\ref{fig:Panel4}). This result is observed across population sizes (Fig.~\ref{fig:Panel3}d).

Of particular interest in reinforcement models is learning speed. Learning speeds can be varied in the model by varying the rate at which weights are accumulated. We systematically investigated the relative speeds of learning for updating network ties and strategies. Previous studies found that varying learning speeds have a noticeable effect on results. Faster network learning speeds tends to promote cooperation \cite{santos2006cooperation,skyrms2000dynamic,pemantle-skyrms2004}. In our model we found varying learning speeds had relatively little effect on resulting behavior. Without errors faster network learning speed tends to produce hybrid or network solutions, closing off the correlated equilibria. With the addition of errors the network solutions disappear and we see mostly the paradoxical convention. When strategy learning is faster agents tend to quickly learn one of the conventional solutions and the network tends to stay more homogenous throughout their evolution (for more detailed discussion of these results see Supplemental Information).

\section{Discussion}

The simulation results support a general conclusion: dynamic networks consistently favor the paradoxical solution over the bourgeois solution across a wide range of parameters. The results for probing the possible payoff space for games of conflict show how frequently simulations produce paradoxical outcomes (Fig.~\ref{fig:Panel3}a). Notice the relative rarity of the bourgeois solution, even for cases where host and visitor attach different values to playing hawk as host versus hawk as visitor (Fig.~\ref{fig:Panel3}b). Only when payoffs are very strongly biased in favor of playing dove when visiting, or against dove when hosting, do we find the bourgeois solution more often than the paradoxical solution (Fig.~\ref{fig:Panel3}c). If there is no bias or any bias in favor of dove when hosting, or against dove when visiting, the paradoxical solution is far more prominent. A common view on the origin of territoriality or norms of ownership takes the evolution and stability of the bourgeois correlated equilibrium to provide a rationale for aggressively defending territories or resolving conflicts over resources in favor of hosts \cite{smith-parker1976,smith1982evolution,skyrms2014evolution,kokko-etal06}. In stark contrast, our results show that a host-guest norm, where the host concedes resources to a visitor, evolves far more readily than any norm of territoriality.

The prevalence of the paradoxical solution occurs because the network solution breaks the symmetry between correlated equilibria. In network solutions hubs are strongly reinforced to play dove as host. If a hub agent finds another dove host to visit then the hub will visit that host more. Once the network connection is strengthened, while hub agents are reinforced to play dove as host by their hawk visitors, they can learn to play hawk as visitor against the other dove host. When this correlated convention is discovered, the rest of the population quickly learns it. Fig.~\ref{fig:Panel3} shows the possible evolutionary routes to the paradoxical solution. It is only when the bias of the payoffs toward hawk-when-hosting and dove-when-visiting is strong enough to overwhelm the influence of the dynamic network that we see the regular evolution of the bourgeois strategy. This is further illustrated by the fact that slower network learning does not lead to the paradoxical solution as frequently as fast network learning (see Supplemental Information).

This result has an important implications for evolutionary accounts of when or where certain conventions may arise. Note that our model represents scarcity of resources indirectly as variation in payoff biases (Fig.~\ref{fig:Panel2}d-e), though we do not consider catastrophic scarcity or interaction outcomes. So, for instance, in the case of territories individuals either are not evicted when conceding to an aggressive opponent, or they can find another territory without severe cost. Also, the model assumes that losses incurred when playing dove against hawks are not overwhelming; individuals do not die or lose all mating opportunities if they concede too many resources. Including scarcity of resources introduces new dynamical elements that may produce different results. If the number of territories is severely limited and a territory is necessary for survival or reproduction, then there is a zero-sum aspect to holding territories that would impact the evolutionary dynamics beyond the game of conflict. The potential significance of these factors would mean that ownership of a territory or control of the resource would confer some intrinsic value and norms of ownership would not play a merely conventional role in avoiding conflict. Our model reveals that in dynamic networks a prevalence territoriality is not due to relatively small differences in efficiency between the bourgeois and paradoxical conventions. These results support the view that territoriality is prevalent because holding a territory has significant value beyond providing an efficient convention for resolving potential conflicts. If there is no scarcity of resources, or no competition over territories, a host-guest convention is much more likely to emerge even when moderately inefficient. Indeed, this may explain why we tend to see host-guest conventions emerge in many domains of human interaction.

Regarding the evolution of convention more generally, Peyton Young defines a convention as a ``pattern of behavior that is customary, expected and self-enforcing'' \cite{young93} and other scholars have identified correlated equilibria as central to conventions \cite{vanderschraaf95}. Our results demonstrate that correlated equilibria can readily emerge in dynamic networks. In this setting, which agent initiates the interaction provides a ready cue for reinforcement learning and sets the stage for correlated equilibria.  Although some accounts of convention involve agents' expectations or beliefs about the behavior of other agents \cite{lewis69,young93,vanderschraaf95}, our model shows that learners with no such beliefs or expectations can establish convention-like behavior (i.e., strategic behavior uniformly adopted from amongst equally viable alternatives that is self-enforcing). Furthermore, many potential conventions in game-theoretic contexts pose problems of symmetry since many conventions are equally efficient. The results presented here show that equally efficient conventions will not evolve with equal probabilities in dynamic networks and thus provides a potential answer to the equilibrium selection problem \cite{samuelson97} in these cases.

In conclusion, our results have two important consequences. First, allowing individuals to avoid aggressive neighbors by choosing their interaction partners through a realistic learning mechanism (reinforcement learning) reveals a novel way to resolve games of conflict by promoting hub-spoke structures that, in turn, break the symmetry between correlated equilibria in a strongly biased way. We know that learning in social networks has profound effects on dynamics for other classes of games and most often the results show that dynamic networks favor cooperation in prisoner's dilemma or stag hunt games \cite{skyrms2000dynamic,pemantle-skyrms2004,santos2006cooperation,rand2011dynamic}. Our study of dynamic networks in games of conflict adds further generality to these results---the network structures that evolve help resolve conflict---and also reveals a surprising tendency to undermine the usual narrative about how the bourgeois convention provides a rationale for norms of ownership or property. Reinforcement learners in dynamic networks have a strong tendency to find the paradoxical solution. Agents only tend to find the bourgeois solution when there are obvious and systematic benefits to territorial behavior: the interaction payoffs for games of conflict need to be biased in such a way that the host values playing hawk much more than a visitor. This entails that insofar ownership and territoriality are probably widespread due to the intrinsic importance of holding resources or the value of owning a territory rather than as a convention for avoiding conflict. Second, simple reinforcement learning in a dynamic network can produce realistic network structures in a way that does not presume a growing network with preferential attachment. Exploring how learning to resolve conflict may interact with preferential attachment to influence the structure of human social networks is an exciting prospect for future research.

\section*{Data Generation and Analysis}

Simulations were written in C++ and run for $10^6$ rounds of play. Data aggregation and network plots were created using Python and R.

When examining asymmetrical payoffs two approaches were used. First, we consider cases where $x_i + y_i = 1$ but varied the $x_i/y_i$ ratio between players 1 and 2. Second, we relaxed the requirement of $x_i + y_i = 1$ and considered points varied at regular payoff intervals.

\section*{Code Availability}
Replication code available on GitHub at \\
\url{https://github.com/riedlc/ConflictAndConvention}.

\section*{Acknowledgements}
This research was supported in part by Army Research Office grant W911NF-14-1-0478, and Office of Naval Research grants N00014-16-1-3005 and N00014-17-1-2542.

\section*{Author contributions statement}
M.F., P.F., R.S., C.R. conceived the experiments, M.F. implemented simulations in C++, M.F. and C.R. conducted the experiment(s), analyzed results, and prepared the figures, P.F., R.S., C.R. wrote the manuscript, M.F., P.F., R.S., C.R. wrote the supplemental information. All authors contributed to all aspects of the project.

\section*{Additional information}
\textbf{Competing financial interests} The authors declare no competing financial interests. 


\bibliographystyle{abbrv}

\end{document}